\documentclass[a4paper,12pt]{article}

\usepackage[utf8x]{inputenc}
\usepackage[T1]{fontenc}
\usepackage[colorlinks=true,linkcolor=black,citecolor=black,urlcolor=blue]{hyperref}
\usepackage{geometry}
\usepackage{titlesec}
\titleformat{\section}
{\bfseries\uppercase}{\thesection.}{1em}{}
\titleformat{\subsection}
{\bfseries}{\thesection.\thesubsection.}{1em}{}

\usepackage{graphicx} 
\usepackage{multirow} 
\usepackage{cite}
\usepackage{breakurl}
\usepackage{indentfirst}
\usepackage{amsmath, amssymb, amsfonts, bm,stmaryrd}
\usepackage{txfonts}
\usepackage{fourier}
\usepackage{enumitem}
\usepackage{xcolor}
\usepackage{enumitem}

\usepackage[sort, numbers]{natbib}

\titlespacing*{\subsection}{0pt}{1.5em}{0.2em}
\titlespacing*{\section}{0pt}{1.5em}{0.2em}

\renewcommand\eqref[1]{Equation~\ref{#1}}

\renewcommand{\thesection}{\arabic{section}}
\renewcommand{\thesubsection}{\arabic{subsection}}

\makeatletter
\renewcommand\@biblabel[1]{#1.}
\makeatother

\newlength{\bibitemsep}
\newlength{\bibparskip}
\let\oldthebibliography\thebibliography
\renewcommand\thebibliography[1]{%
  \oldthebibliography{#1}%
  \setlength{\parskip}{\bibitemsep}%
  \setlength{\itemsep}{\bibparskip}%
}

\newcommand{\YearConf}{2024}

\newcommand{\LogoConf}{logo_IN24.jpg}
\newcommand{\CopyrightConf}{Permission is granted for the reproduction of a fractional part of this paper published in the Proceedings of INTER-NOISE \YearConf ~ \underline{provided permission is obtained} from the author(s) \underline{and credit is given} to the author(s) and these proceedings.}

\usepackage{fancyhdr}
  \fancypagestyle{firststyle}
{   \fancyhf{}
   \fancyfoot[C]{\scriptsize \CopyrightConf}
}
\begin{document}
\thispagestyle{firststyle}

\begin{center}
	\includegraphics[width=2in]{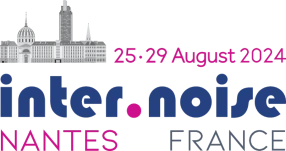}
\end{center}
\vskip.5cm

\begin{flushleft}
\fontsize{16}{20}\selectfont\bfseries
\color{black}Towards better visualizations of urban sound environments: insights from interviews \\

\end{flushleft}
\vskip1cm

\renewcommand\baselinestretch{1}
\begin{flushleft}


Modan Tailleur\footnote{modan.tailleur@ls2n.fr}\\
Nantes Université, École Centrale Nantes, CNRS, LS2N\\
F-44000 Nantes, France\\

\vskip.5cm
Pierre Aumond\\
Univ Gustave Eiffel, CEREMA, UMRAE\\
F-44344 Bouguenais, France\\

\vskip.5cm
Vincent Tourre\\
Nantes Université, École Centrale Nantes, CNRS, AAU, UMR 1563\\
F-44000 Nantes, France\\

\vskip.5cm
Mathieu Lagrange\\
Nantes Université, École Centrale Nantes, CNRS, LS2N\\
F-44000 Nantes, France\\

\end{flushleft}
\textbf{\centerline{ABSTRACT}}\\
\textit{Urban noise maps and noise visualizations traditionally provide macroscopic representations of noise levels across cities. However, those representations fail at accurately gauging the sound perception associated with these sound environments, as perception highly depends on the sound sources involved. This paper aims at analyzing the need for the representations of sound sources, by identifying the urban stakeholders for whom such representations are assumed to be of importance. Through spoken interviews with various urban stakeholders, we have gained insight into current practices, the strengths and weaknesses of existing tools and the relevance of incorporating sound sources into existing urban sound environment representations. Three distinct use of sound source representations emerged in this study: 1) noise-related complaints for industrials and specialized citizens, 2) soundscape quality assessment for citizens, and 3) guidance for urban planners. Findings also reveal diverse perspectives for the use of visualizations, which should use indicators adapted to the target audience, and enable data accessibility. 
}
\section{Introduction}




Noise is a major problem in urban areas, leading to the exploration of various methods to visualize its impact on city life. Directive 2002/49/EC of the European Union \cite{european_commission_european_2022} mandates noise level mappings in urban areas with over 100,000 residents and near major transportation hubs. These mappings are derived from computer modeling that integrates descriptive data on topography, meteorology,... and noise sources, obtained from multiple stakeholders. They show several acoustic indices corresponding to noise levels weighted throughout different time periods throughout the day.


In addition to noise maps, measurement campaigns are sometimes organized to gain a more comprehensive understanding of urban noise \cite{aumond_modeling_2017, torija_application_2013, nilsson_acoustic_2007, can_cense_2021, mietlicki_innovative_2015, farres_barcelona_2015}. \noindent Organizations responsible for monitoring environmental noise, such as Bruitparif \footnote{https://www.bruitparif.fr/} and Acoucité \footnote{https://www.acoucite.org/}, have consequently introduced various visualization platforms to provide the general public and technical services of cities access to those measurements. These platforms enable users to conduct a more detailed temporal analysis and explore additional noise indices, facilitating deeper insights into urban soundscapes. 

While noise maps and aforementioned monitoring platforms developed by noise observatories sometimes offer more soundscape-oriented acoustic indices, such as the \textit{Harmonica} index \cite{mietlicki2015new}, they usually don't provide in-depth perceptual insights. Furthermore, even if those platforms are sometimes specialized for one type of sound source, such as the \textit{Survol} platform of Bruitparif dedicated to aircraft noise, they do not provide a comprehensive analysis of the impact of the various sound sources on the soundscape.  However, differentiating the different sound sources of an environment is crucial for understanding the perception of noise \cite{lavandier_contribution_2006}. For instance, a high noise level could originate from various sources such as voices, construction work, or road traffic, each perceived differently.


Recent research has shown visualizations of the presence of multiple sound sources in urban environments \cite{lavandier2016urban, moreno2022visual, lassowebsite}. If representing the sound of cities more accurately by incorporating multi-source information appears to be valuable, it also introduces another level of complexity to the visualizations. Therefore, this paper aims to understand whether incorporating multiple sound sources information could enhance the analysis of the different city stakeholders that are currently using noise visualizations. Additionally, it aims to comprehend which stakeholders currently utilize noise visualizations and to what extent the existing visual representations meet their needs. Qualitative interviews are conducted with various stakeholders, including non-profit environmental organizations responsible for noise monitoring, city technical services, officials, and non-specialists. This diverse panel reflects all stakeholders involved in creating, consulting, and making decisions based on noise-related visualizations. Following evaluation methodologies previously considered in the literature \cite{kinkeldey2015evaluating, parsons2021understanding},  we believe that these interviews provide valuable insight into the perspectives and preferences of stakeholders. The findings notably uncover three different uses of sound source representations: 1) noise-related complaints for industrials and specialized citizens, 2) soundscape quality assessment for citizens, and 3) guidance for urban planners.

Section \ref{sec:Methods} offers details on the methodological framework utilized during the interviews. Section \ref{sec:Results} presents an extensive analysis of the diverse use cases that emerged from the interviews. Additional complementary topics are explored in Section \ref{sec:Discussions}.

\section{Interviews}\label{sec:Methods}

\subsection{Selection of participants  \label{Sec:MethodParticipant}}

11 diverse interviews have been conducted with 8 distinct groups of city stakeholders, providing a comprehensive perspective on soundscape-related considerations. The interviewees are categorized into various roles and affiliations.

\vspace{12pt}

\noindent Non-profit environmental organization responsible for monitoring environmental noise (OBS):
\begin{itemize}
\item OBS1:
    \begin{itemize}
        \item OBS1.1: director of the association.
        \item OBS1.2: technician working with data processing and storage.
        \item OBS1.3: technician working on sensors.
    \end{itemize}
\item OBS2:
    \begin{itemize}
        \item OBS2.1: director of the association.
        \item OBS2.1: technical director.
        \item OBS3.1: PhD student working on source recognition algorithms.
        \item OBS3.4: project manager in noise and transport.
    \end{itemize}
\end{itemize}

\noindent Technical services of cities (TS):
\begin{itemize}
\item TS1: Officier in charge of ecological transition, air quality, and noise pollution control for a french metropolis.
\item TS2: Officier in charge of the Environment, Ecological Transition and Sustainable Development for a french city.
\end{itemize}

\noindent City officials (CO):
\begin{itemize}
\item CO1: The official of a french city on the questions of land use. 
\end{itemize}

\noindent Urbanist (URB)
\begin{itemize}
\item URB1: Urban planner specialized in territorial planning.
\end{itemize}

\noindent Project Owner's Assistant (POA)
\begin{itemize}
\item POA1: project owner's assistant. Accompanies real estate operators in their development.
\end{itemize}

\noindent Acoustician (AC):
\begin{itemize}
\item OC1: Acoustical engineering consultant.
\end{itemize}

\noindent Geomatician (GEOM)
\begin{itemize}
\item GEOM1: Geomatician, in close relation with acoustician researchers.
\end{itemize}

\noindent Non specialists (NS):
\begin{itemize}
\item NS1: Front-end developer. 
\item NS2: Orthophonist. 
\end{itemize}

\vspace{12pt}

Relevant findings identified as relevant for specific stakeholder groups are denoted in parentheses using corresponding group shortcuts (e.g., OBS) in the subsequent sections.

\subsection{Structure of Interview}

Several interviews are conducted with one or more city stakeholders. Interviews are composed of several parts, each described in the following sections.

\subsubsection{Gathering information about past experience (10 minutes)}

The interviewees provide information regarding the scope of their activity, and their potential responsibilities in noise management. Additional information is collected, including the duration of their professional experience, interactions with other stakeholders, current practices, potential use of source recognition tools in their profession, and other relevant factors. 



\subsubsection{Demonstration of existing platforms (5 minutes)}

\begin{figure*}[!t]
\begin{center}
\includegraphics[width=\linewidth]{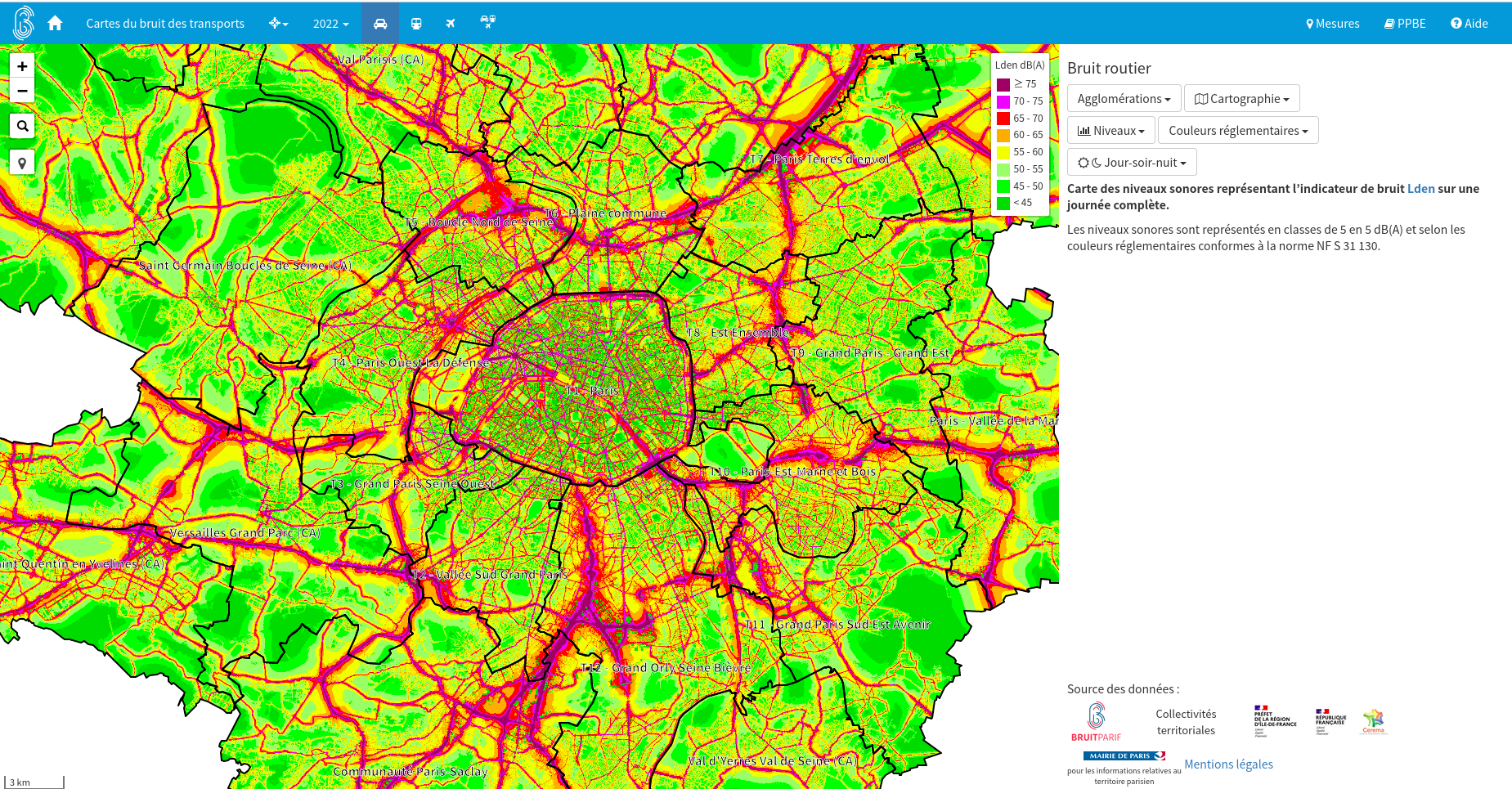}
\end{center}
\caption{Noise map of the city of Lorient\cite{bruitparifwebsite}}
\label{fig:carte_1}
\end{figure*}

\begin{figure*}[!t]
\begin{center}
\includegraphics[width=\linewidth]{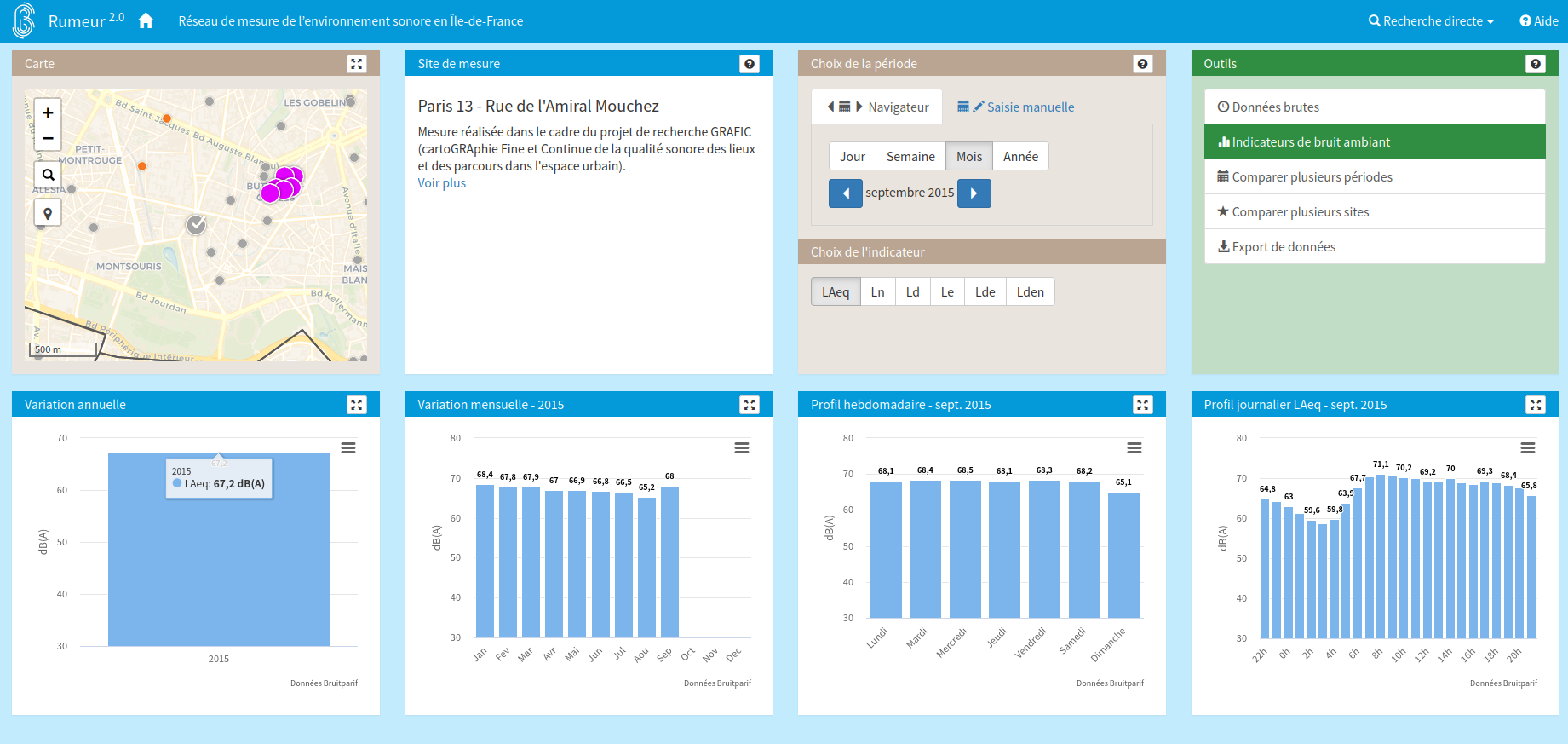}
\end{center}
\caption{\textit{Rumeur} platform\cite{bruitparifwebsite}}
\label{fig:rumeur_1}
\end{figure*}

The participants are presented with a demonstration of two specific visualizations developed by Bruitparif: a noise map\footnote{https://carto.bruitparif.fr/}(see figure \ref{fig:carte_1}), and \textit{Rumeur}\footnote{https://rumeur.bruitparif.fr/}(see figure \ref{fig:rumeur_1}). Bruitparif is a noise observatory, working on sound environment characterization, participating in scientific studies, collaborating with stakeholders to integrate noise considerations into public policies, and sharing information to raise public awareness about soundscapes in the Île-de-France region. These tools are among the most widely and advanced used resources provided by noise observatories in France.



\begin{figure*}[!t]
\begin{center}
\includegraphics[width=0.5\linewidth]{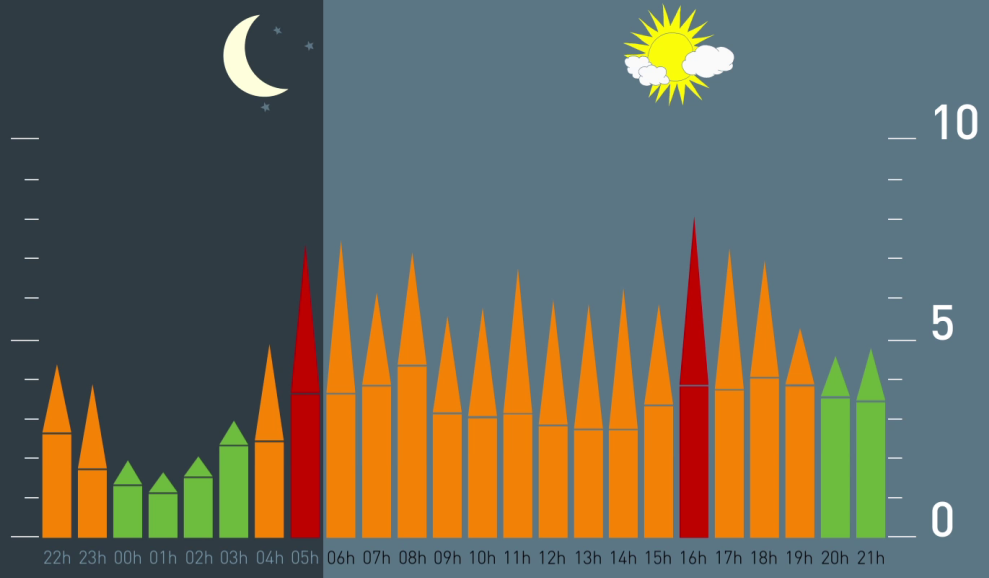}
\end{center}
\caption{Example of \textit{Harmonica} index throughout the day \cite{bruitparifwebsite}}
\label{fig:harmonica example}
\end{figure*}

\textit{Rumeur} serves as a platform for noise monitoring, encompassing data from multiple sensor networks: some from completed projects, and others currently in operation. The interviewees are provided with a comprehensive overview of the platform's capabilities, including site selection, consultation of noise indicators, comparison of different periods or sites, and visualization and playback of sound events.  One of the showcased indicators is the \textit{Harmonica} index \cite{mietlicki2015new}, a graphical representation offering insights on two major components that impact the sound environment: the level of the ambient background noise, and the level of sound events that emerge from this background noise (see figure \ref{fig:harmonica example}).

Subsequently, the interviewees are asked about their current utilization of noise maps or platforms such as \textit{Rumeur} in their professional activities, and their willingness to consider employing such tools in the future.

In the following sections, Rumeur-like platforms are referred to as OBS-platforms, and noise maps are referred to as NM-platforms.

\subsubsection{Strengths and Weaknesses of existing platforms (10 minutes)}

The interviewees are asked to share their insights on the strength, weaknesses, and potential enhancements to NM-platforms and OBS-platforms, considering each visualization independently. Participants are then invited to provide their perspectives on the potential user groups for these visualizations. 

\subsubsection{Visualization of the presence of multiple sound sources (5 minutes)}

After the introduction of PANN-1/3oct\cite{tailleur_spectral_2023}, a sound classification algorithm based on PANNs\cite{kong_panns_2020}, participants are asked about their preferences regarding visualization of specific sound sources, namely traffic, voices, and birds. Specifically, they are invited to express whether they find it beneficial to visualize sound sources data on NM-platform or OBS-platform, and to elaborate on how they would envision this representation. Furthermore, participants are encouraged to share their opinions on the adequacy of the three identified sources (traffic, voices, and birds) and whether additional categories or refinements could be useful.

In addition to exploring participant preferences, interviewees are asked about other potential stakeholders who might express interest in having sound sources presented in similar visualizations. 

\subsubsection{Ideas for Visualization and Analysis  (10 minutes)}

\begin{figure*}[!t]
\begin{center}
\includegraphics[width=0.75\linewidth]{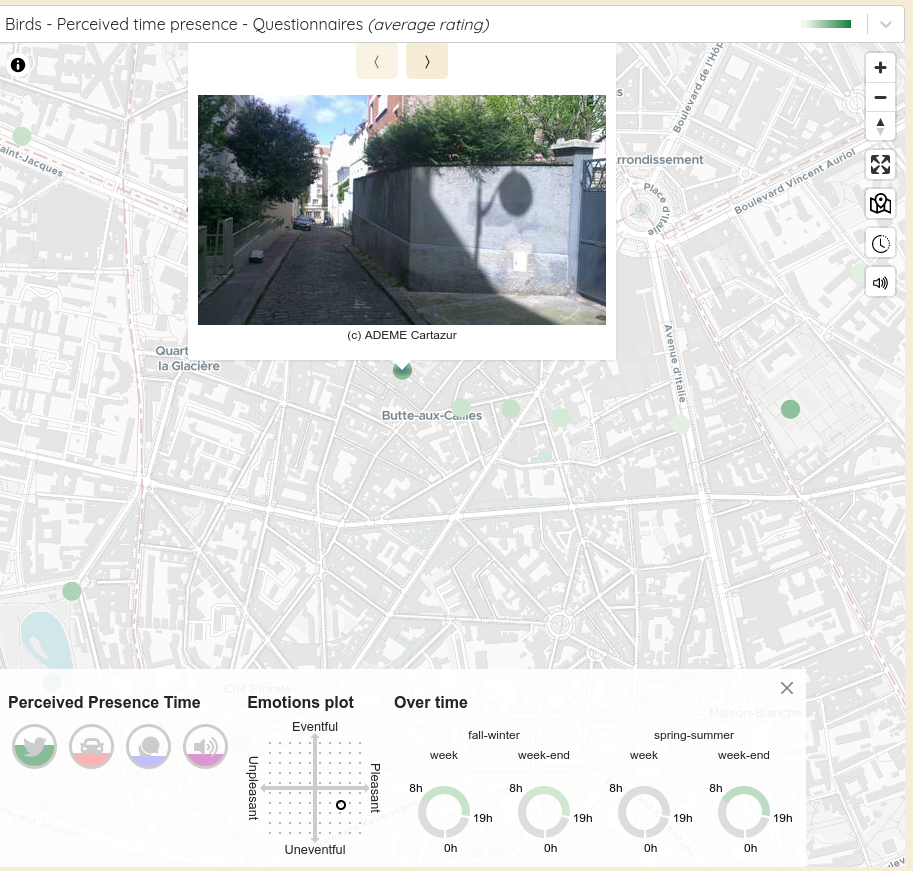}
\end{center}
\caption{\textit{Lasso} platform\cite{lassowebsite}}
\label{fig:lasso}
\end{figure*}

A demonstration of the \textit{Lasso} platform\footnote{https://universite-gustave-eiffel.github.io/lasso/} is provided. \textit{Lasso} is a web-based platform that offers a range of spatio-temporal datasets related to soundscapes, including distinctive functionalities. Notably, within the \textit{Lasso} platform, participants are introduced to a scale ranging from 0 to 10, illustrating the time of presence of traffic, voices, and birds. In addition, a graph representing the eventfulness and pleasantness of the sound is displayed on the platform. \textit{Lasso} also presents two different types of maps, one with points of measure and the other with interpolated data. An example of the \textit{Lasso} platform on points of measure is shown in figure \ref{fig:lasso}. Platforms similar to \textit{Lasso}, which provide multi-source visualizations of the soundscape, are referred to as MS-platforms in the subsequent sections.

Depending on the context of the conversation, supplementary examples are presented that showcase various visualization and analysis techniques.

\subsubsection{Exploring Additional Ideas (10 minutes)}

Discussions are conducted to gather insights and ideas from the interviewees. Topics include the identification of the actors that would be the most interested and the type of visualization that would be the most suited for them, but also any other suggestions or perspectives that align with their professional practices or the one of other stakeholders.

\section{Results}\label{sec:Results}

As highlighted by OBS2.2, \textit{"NM-platforms and OBS-platforms are two things that no one really manages to merge. They are complementary because they are developed in different ways."}. Informed by this insight, we strive to discern distinctive use cases for each of these two separate methods of sound representation. Additionally, we aspire to investigate a third use case centered around urban planners with a more methodological focus.

\subsection{Noise-Related Complaints}

In the realm of sound maps, a notable use case revolves around addressing noise-related complaints, offering particular utility to citizens and industrial stakeholders. This use case is particularly centered on citizens who are already familiar with the technicalities of acoustic indices and who live in areas affected by noise, such as those close to airports. The role of source recognition would be to effectively distinguish various sound sources and assess their respective contributions to the overall noise levels. With such discrimination between the different sounds, industrial stakeholders would also have a better understanding on whether the noise in an area is attributable to their activity, or to the one of other stakeholders.

Several key requirements and suggestions surface for noise related complaints visualization:
\begin{itemize}
\item There is a need to represent detailed data collected at precise time intervals and locations (OBS, TS, CO). A visual map may even not be a prerequisite (NS), as the interest is focused at a very specific location. 

\item Displaying multiple sound sources on one visualization may not be crucial, as the primary objective is to identify the most problematic sound source and to analyze it (OBS). It may even be beneficial to have a separate OBS-platform for each distinct noise source (OBS), considering that each source has its specific indices and analyses. 

\item The visualization should also encompass precise noise indicators (TS) to accurately represent the noise. Simultaneously, the visualizations should include simple indicators to facilitate understanding, especially for citizens. Notably, the \textit{Harmonica} index and the audio playback tool for listening to the audio at a specific location and a specific time are considered particularly interesting (NS). 
\end{itemize}

\subsection{Soundscape Exploration}

A second use case would be to provide a broader representation of the soundscape, at the scale of the city. The primary target audience encompasses the general population, especially those considering residing in a new neighborhood, or seeking information on the sonic characteristics of various areas (OBS, NS). It can also serve pedagogical purposes, by demonstrating the impact of human activities (TS). Real estate agents can leverage this tool to support the sale of buildings, aiding citizens in making informed decisions about where they want to live (NS, POA). The role of source recognition in soundscape consultation is to visually articulate the distribution and evolution of sound sources within a specific environment at a given time period.

Several requirements and suggestions emerged from the interviews for general soundscape consultation:

\begin{itemize}
\item Data should be represented on interpolated maps rather than points of measurement, to fulfill the need for visual clarity (NS, TS, CO). Some users may expect sound sources to match their experiences of annoyance or pleasantness at a specific location. However, concerns have been raised about the precision of interpolated maps, as the interpolation can make the interpretation between points of measurement extremely misleading (GEOM, TS, CO).  Potential solutions, such as marking specifically the points of measurement on the map, or presenting noise data over substantial areas (e.g. aggregating by neighborhoods), may address this challenge (CO, GEOM).

\item To cater to the diverse audience, it is essential to use indicators that are accessible and meaningful to the general public (NS, CO, TS). As emphasized by CO1, \textit{"residents do not recognize themselves in the noise maps presented as they are today. There is really a strong sense of a gap between the sound map and the citizen's perception."}. The graph of pleasantness and eventfulness showcased on \textit{Lasso} help bridging this gap by making the information more relatable to residents (NS, CO, TS). Although data for soundscape consultation can be less precise, it still needs to be meaningful. In this context, the option of listening back to the audio of the area of interest adds an enriching dimension to the user experience (NS, CO, TS).  

\item If more complex visualizations or indices are introduced, the platform should incrementally increase in complexity rather than presenting them directly (TS, NS). This approach ensures that users can engage with the information at a pace that suits their level of familiarity and expertise.

\end{itemize}

\subsection{Guidance for urban planners}

In the realm of urban planning, a unique use case emerges. Combining various information domains, including air pollution, soil pollution, and topology, urban planners face the challenge of gathering and presenting extensive data (URB, POA). They present those very diverse data to different stakeholders, such as city officials. Introducing both a NM-platform and an OBS-platform as representations of noise in this context might lead to presentations that would be perceived as overwhelming. Consequently, a balanced integration of both representations functionalities is likely desired. Their approach to sound representation is also intricately tied to the project's specific context, as it varies based on project dependencies.

Several essential prerequisites emerge to assist urban planners in developing new visualizations:
\begin{itemize}
\item In building visualizations useful for urban planners, precision becomes paramount, as noise levels need to be assessed at an extremely localized scale (URB, POA). Urban planners are typically focused on specific neighborhoods or streets, rather than citywide considerations, given the limited options for construction sites. The visualization needs to be centered around the building of interest, to evaluate potential noise-generating areas nearby. Consequently, a detailed map of the neighborhood, rather than a generalized city map, proves more pertinent for their purposes (TS, URB, POA). As underscored by POA1: \textit{"When dealing with a program or operation at the scale of a street, the information needs to be highly localized because the situation can vary from one street to another, from one plot to another. In fact, it's this quality of information that we are looking for: ultra-localization."}. 

\item Although urban planners may find value in representations such as NM-platforms and OBS-platforms, they may not necessarily use them directly in their reports in their current state. Rather, they prefer direct access to raw data, enabling them to create personalized representations tailored to the context of each project (URB). As they often provide some cross-representations between sound and other topics, they would need to have coherent representation types between their different visualizations. Instead of providing them pre-defined platforms, providing a variety of visualization ideas and examples can assist urban planners in exploring methodologies, visuals, and analyses that are both aesthetically appealing and scientifically meaningful. As URB1 explains, \textit{"To assist the urban planner is not necessarily to define, in their place, the best possible representations; it is primarily to share the data to allow them to create their own representations."}.

\item In contrast to other use cases, urban planners place less emphasis on the need for interpretable information for the general public. Indices such as the Harmonica index and the \textit{Lasso} graph on pleasantness and eventfulness are considered too vague for their purposes. Similarly, features like listening back to audio recordings are deemed less useful in this context. Urban planners prioritize precise measurements of noise levels to inform their decision-making processes (URB). To assess the perceptual quality and emotions engendered by the sound of the city, urban planners would prefer to gather on-the-spot information through questionnaires from inhabitants to obtain a more authentic understanding of the sound perception (URB).
\end{itemize}

\section{Discussion}\label{sec:Discussions}

\subsection{On noise indices}

Noise indices currently proposed in NM-platforms and OBS-platforms, such as LAeq, Ln, and Lden, present accessibility challenges for non-acousticians and non-specialists (OBS, ST, NS, AC). Insights from interviews revealed two distinct solutions to address this challenge.

The first solution is the promotion of public education concerning existing sound indicators incorporated into city representations. Some stakeholders emphasized the pressing need for pedagogy surrounding urban soundscapes, directed towards both the general public and city officials (OBS). Discussed strategies involve incorporating features that allow individuals to listen to audio while simultaneously receiving relevant sound indicators, providing them with an intuitive understanding of the noise measurements. Another feature example would be facilitating comparisons between different locations and time periods, akin to functionalities present in platforms like \textit{Rumeur}. 

A second strategy involves the introduction of new indices specifically tailored for the general public. Examples of these indices have received positive feedback from the interviewees. Although the \textit{Harmonica} index is widely appreciated (OBS, TS, CO), concerns have been voiced regarding its complexity, as it is still perceived as somewhat expert-oriented (NS). The \textit{Lasso} diagram representing pleasantness and eventfulness was also considered highly relevant and accessible for the general public (NS, TS, AC).

\subsection{On data interpretation}

To instill confidence in officials regarding the potential interpretation of data by citizens, the importance of data transparency and raw data access cannot be overstated (ST). A clear description of the methods of calculation or measurement is essential, and the visualization should effectively communicate their limitations. This may include measures like restricting the zoom level (GEOM, TS), aggregating data at the neighborhood scale (CO, NS, GEOM), or displaying only measurement points (ST). To prevent misinterpretation, there also is a suggestion to present pre-analysis results instead of raw data, aiding individuals in understanding the visualization (CO, ST). 

Given that sound source predictions lack standardization, concerns about their interpretation have also been raised (TS). A proposed solution involves providing audio examples for key sound source presence levels directly within the visualization, or simply allowing playback of audio. 

\section{Conclusion}


Two primary forms of noise representations are currently provided by noise observatories: noise maps, and in-depth visualizations of measurement data. Interviews with various city stakeholders have been conducted to understand which benefit the most from those representations of noise. This paper also aimed to assess whether these visualizations could be enhanced by incorporating information about sound sources, and which types of stakeholders would find it most valuable.  Throughout our investigation, three distinct applications emerged:


\begin{itemize}
    \item \textbf{Noise Complaints}: It leans toward a visualization approach resembling platforms like \textit{Rumeur}, aiming both the general public and industrial stakeholders. For sound sources detection, the platform would probably be centered around one key noise source.

    \item \textbf{Soundscape Exploration}: Aligned more closely with sound mapping, it is primarily intended for the general public, emphasizing the spatial diversity, distribution, and evolution of sound sources within a city. 

    \item \textbf{Guidance for Urban Planners}: Providing recommendations for urban planners, would guide them on how to represent the soundscape effectively, choose relevant indices, address potential misinterpretations, and potentially help them integrate sound representations with other environmental data, such as air pollution.
\end{itemize}

In all these applications, it was found that the choice of noise indices should be carefully considered, either by promoting current indices to the target city stakeholder, or by designing new noise indices specifically tailored for them. To avoid misinterpretation of the data, a clear emphasis needs to be placed on data transparency.




\section{Acknowledgements}

The authors would like to express their gratitude to all the individuals and organizations who participated in the interviews conducted for this research. This work has been funded by the AIby4 project (Centrale Nantes and Project ANR-20-THIA-0011).

\bibliographystyle{unsrt}
\bibliography{references}

\newpage \pagebreak \cleardoublepage


\end{document}